%% file: codehealth.tex
\documentclass[sigconf]{acmart}

\usepackage{booktabs} % For formal tables

\usepackage{balance}
\usepackage{multirow}

\usepackage{graphicx}

\setcopyright{none}
\begin{document}
\title[Code Red: The Business Impact of Code Quality]{Code Red: The Business Impact of Code Quality\\-- A Quantitative Study of 39 Proprietary Production Codebases}
%\subtitle{}

\author{Adam Tornhill}
\orcid{XXX}
\affiliation{%
  \institution{CodeScene}
  \city{Malmö}
  \country{Sweden}
}
\email{adam.tornhill@codescene.com}

\author{Markus Borg}
\orcid{XXX}
\affiliation{%
  \institution{RISE Research Institutes of Sweden\\Lund University}
  \city{Lund}
  \country{Sweden}
}
\email{markus.borg@ri.se}

% The default list of authors is too long for headers.
\renewcommand{\shortauthors}{A. Tornhill and M. Borg}

\begin{abstract}
Code quality remains an abstract concept that fails to get traction at the business level. Consequently, software companies keep trading code quality for time-to-market and new features. The resulting technical debt is estimated to waste up to 42\% of developers' time. At the same time, there is a global shortage of software developers, meaning that developer productivity is key to software businesses.
Our overall mission is to make code quality a business concern, not just a technical aspect. Our first goal is to understand how code quality impacts 1) the number of reported defects, 2) the time to resolve issues, and 3) the predictability of resolving issues on time.
We analyze 39 proprietary production codebases from a variety of domains using the CodeScene tool based on a combination of source code analysis, version-control mining, and issue information from Jira. %This combination allows us to track the development time and issue resolutions down to the file level. 
By analyzing activity in 30,737 files, we find that low quality code contains 15 times more defects than high quality code. Furthermore, resolving issues in low quality code takes on average 124\% more time in development. Finally, we report that issue resolutions in low quality code involve higher uncertainty manifested as 9 times longer maximum cycle times.
This study provides evidence that code quality cannot be dismissed as a technical concern. With 15 times fewer defects, twice the development speed, and substantially more predictable issue resolution times, the business advantage of high quality code should be unmistakably clear.
\end{abstract}

%\copyrightyear{2022} 
%\acmYear{2022} 
%\acmConference[TechDebt '22]{International Conference on Technical Debt}{May 22--23, 2022}{Pittsburgh, USA}
%\acmBooktitle{Proceedings of the 3rd ACM SIGSOFT International Workshop on Machine Learning Techniques for Software Quality Evaluation (MaLTeSQuE '19), August 27, 2019, Tallinn, Estonia}
%\acmPrice{15.00}
%\acmDOI{10.1145/3340482.3342742}
%\acmISBN{978-1-4503-6855-1/19/08}

%
% The code below should be generated by the tool at
% http://dl.acm.org/ccs.cfm
% Please copy and paste the code instead of the example below.
%

%\ccsdesc[500]{Software and its engineering~Software verification and validation}
%\ccsdesc[500]{Computing methodologies~Simulation evaluation}
%\ccsdesc[500]{Software and its engineering~Search-based software engineering}
%\ccsdesc[500]{Computing methodologies~Object detection}
%\ccsdesc[300]{Computer systems organization~Embedded and cyber-physical systems}


\maketitle

\input{body.tex}

\balance
\bibliographystyle{ACM-Reference-Format}
\bibliography{codescene}

\end{document}

%% file: body.tex
\section{Introduction} \label{sec:intro}
Efficient software development is a competitive advantage that allows companies to maintain a short time-to-market~\cite{ebert2020technology}. To succeed, companies need to invest in their software to ensure efficient and fast marketplace results~\cite{lesser2016leading}. Adding to that challenge, software companies are also facing a challenge in that there is a global shortage of software developers~\cite{breaux20212021}; demand substantially out-weights supply. Moreover, several analyses forecast that the shortage of software developers will only be exacerbated as the digitalization of society continues~\cite{ozkaya2021developer,hyrynsalmi2021war,breaux20212021,news24wire,borg2020illuminating}. At the same time as the software industry is struggling with recruiting enough talent, research indicates that up to 42\% of developers' time is wasted dealing with Technical Debt (TD) where the cost of subpar code alone comes to \$85 billion annually~\cite{stripe}. This implies that there is an untapped potential in software projects if the code quality is improved and TD paid down.

Unfortunately, the software industry often moves in the opposite direction: research finds that developers are frequently forced to introduce new TD~\cite{besker2019software} as companies keep trading code quality for time to market and new features~\cite{giardino2015software}. This is a decision-making bias known as \textit{hyperbolic discounting}, i.e., humans make choices today that their future selves would prefer not to have made~\cite{laibson1997golden}.

One reason for the hyperbolic discounting of code quality is that the business impact of code quality remains vague~\cite{kruchten2019managing}. This was painfully visible in a study of 15 large software organizations where the benefits of paying down TD is not always clear to managers, and consequently some managers would not grant the necessary budget, nor priorities, for  refactoring~\cite{martini2018technical}. The lack of clear and quantifiable benefits makes it hard to build a business case for code quality and, hence, easy to trade short-term wins for long-term sustainability and software maintenance; there are no standard Key Performance Indicators (KPIs) for code quality that are relevant to companies in the way that financial KPIs are~\cite{baggen2012standardized}. Consequently, in a study involving 1,831 participants, only 10\% reported that their business manager was actively managing TD~\cite{ernst2015measure}. To make a change, we conclude that TD needs visibility throughout the whole organization, which includes business managers, not just developers and architects.

Second, software organizations lack a way of tracking the time wasted on TD with sufficient accuracy~\cite{guo2011tracking}. Further, enforcing such detailed time tracking could be perceived as exercising too much control over employees~\cite{reik2012lean} as well as adding too much administrative burden ~\cite{besker2019software}. In consequence, Besker \textit{et al.} report from a study surveying 43 developers that ``none of the interviewed companies had a clear strategy on how to track and address the wasted time''~\cite{besker2019software}. This means that while the overall development costs (e.g. staffing) are known, there is a need for better ways of mapping those overall costs to the additional time spent working on production code of various quality, i.e., the TD interest.

In this paper, we report how code quality impacts development time in 39 proprietary production codebases. We do that by considering CodeScene's Code Health metric\footnote{\url{https://codescene.com/code-health}} as a proxy for code quality. Moreover, we explore an algorithm to capture the development time at the file level, we refer to this fine-granular metric as Time-in-Development. Hence, this study combines source code metrics with repository mining and issue information from Jira.

Our results clearly show that developers spend more time resolving issues in low-quality source code. We find that change implementation, i.e., either defect resolution or new feature development, in low-quality code is afflicted by higher uncertainty -- the variance of the Time-in-Development is substantially higher when code quality is low. For changes of comparable complexity, the change implementation time is on average more than 2 times longer in low-quality code. Moreover, we have observed several examples for which the development time is an order of magnitude longer in low-quality code.

We strongly voice that maintaining a high code quality must be considered a fundamental business goal. This study contributes a quantification of the negative consequences of low code quality, i.e., figures in development time that can be directly translated into costs. Our findings help raising the awareness of the excess waste in development time as well as the unpredictability correlated with changes to low-quality code. Unpredictability in particular is likely to cause stress to an organization since commitments and deadlines are put at risk. In addition, the new perspective on code quality that we present helps organizations put a value on refactoring and code quality improvements.

The rest of this paper is organized as follows. Section~\ref{sec:bg} presents the theory behind the code quality metric that is at the core of our study, i.e., the Code Health measure provided by CodeScene. Section~\ref{sec:rw} presents an overview of the relevant research on code quality, TD, and Time-in-Development, and positions our contribution in the light of the business impact. In section~\ref{sec:method}, we outline the design of our study. Section~\ref{sec:results} presents our results and we discuss them from a software business perspective in section~\ref{sec:disc}. Finally, we conclude the paper in section~\ref{sec:conc} and propose directions for future work.

\section{Background} \label{sec:bg}
As pointed out in the introduction, there is no industry-wide standard for code quality. Furthermore, there is no established way to track the time wasted on low code quality and TD. This section defines the metrics for code quality and Time-in-Development used in this study.

\subsection{Measure Code Quality by Code Health} \label{sec:codehealth}
There are numerous studies on code quality metrics~\cite{riaz2009systematic,baggen2012standardized}. Although primarily used to communicate how comprehensive a software solution is or how much work a development task would require, Lines of Code (LoC) metrics are often involved also in code quality metrics~\cite{alpernas2020wonderful}. An underlying assumption is that the higher LoC a file has file, the more complex it is. Also within the scope of our research, we assume that making a change in a large file would typically require more effort than in a smaller file. To verify that assumption, we prepared our research by investigating if LoC could be used to predict issue resolution times in our data set (further described in Section~\ref{sec:data}). We found a low Pearson correlation (r=0.13) between added LoC and the average issue resolution time in corresponding files. One reasonable explanation for the low correlation is that the naive LoC cannot distinguish between large data files without application logic, e.g., configuration files and enum files, versus truly complex units~\cite{tornhill2015your}. 

We also note that typing is not the bottleneck in programming; in Robert Glass's words, ``the dominant maintenance activity'' is understanding the existing code~\cite{glass2002facts}. Recent research indicates that, on average, developers spend about 60\% of their time on program comprehension activities~\cite{xia2017measuring}. In this light, the quality of the existing code plays a significant role: a study by Politowski \textit{et al.} shows that code containing the anti-pattern Spaghetti Code is 39.5\% more time consuming to understand than code without this particular code smell~\cite{politowski2020large}. All together, we identified the need for a more comprehensive metric that includes code smell detection -- thus we turned our attention to CodeScene's Code Health.

\begin{figure}
    \centering
    \includegraphics[width=0.5\textwidth]{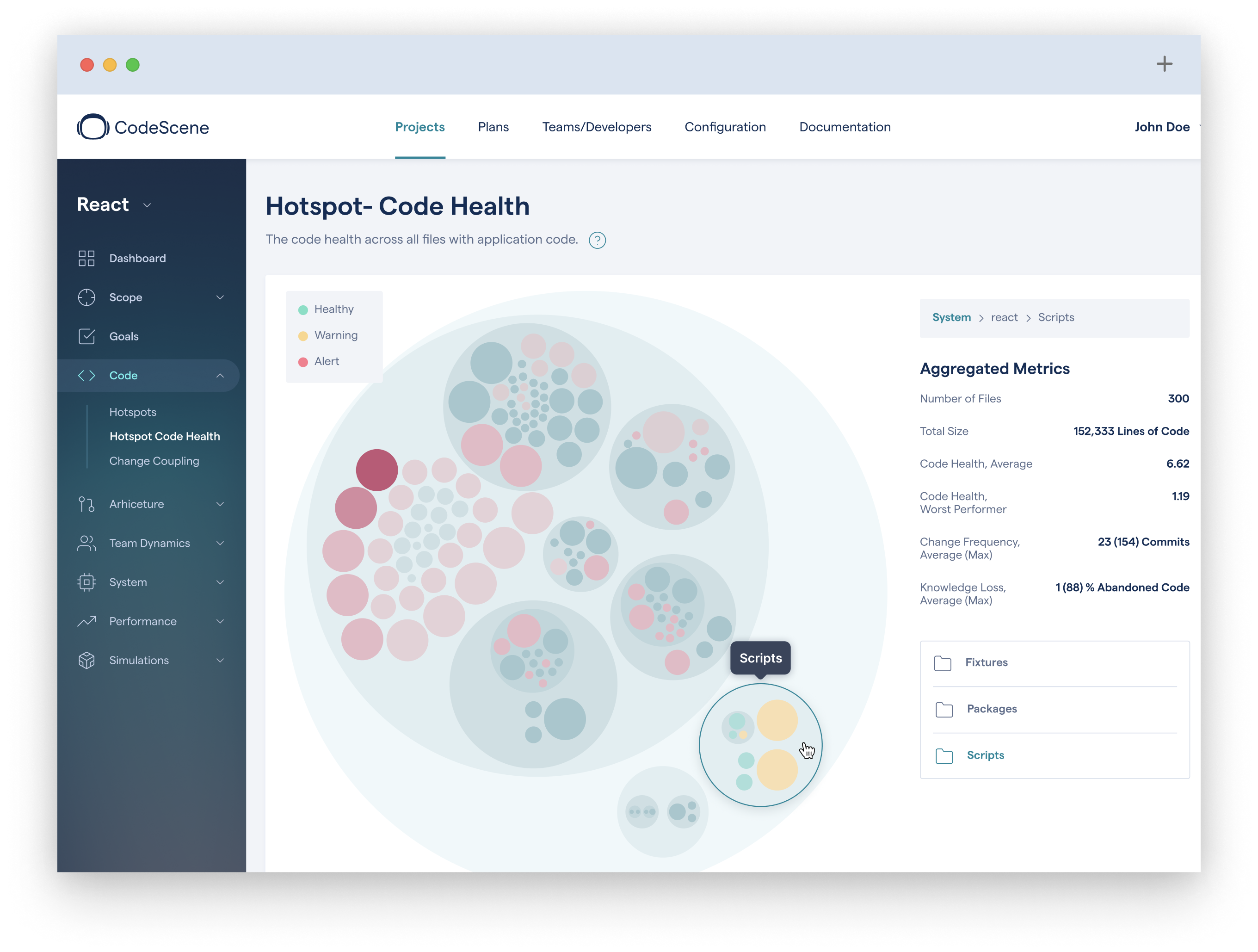}
    \caption{Example Code Health visualization of a codebase.}
    \label{fig:codehealthexample}
\end{figure}

Code Health is a proprietary metric that is automatically calculated in the CodeScene tool. The metric is a numeric score ranging from 10.0 (healthy code of high quality) down to 1.0 (unhealthy code of low quality). Based on the Code Health scores, CodeScene classifies each module as either healthy, warning, or alert and visualizes the codebase accordingly. Healthy corresponds to high code quality (green), whereas warning (yellow) and alert (red) represent code of lower quality (see Figure~\ref{fig:codehealthexample}). The cut-off point for healthy code is 8.0, for alert 4.0, and the range in between represents the warning space. CodeScene reports that the cut-off points were decided by their internal team via a baseline library of hand-scored code examples. CodeScene is available for free in the community and research editions~\cite{tornhill2018prioritize}.

The Code Health metric is modelled around the idea that the most promising approach to measuring code complexity is by identifying and measuring specific attributes of complexity~\cite{fenton1994software}. As such, Code Health incorporates elements from the generic ISO~5055 maintainability section, and complements them with design-level code smells\footnote{\url{https://docs.enterprise.codescene.io/versions/5.1.0/guides/technical/code-health.html}}. Examples of included design smells are God Class, God Methods, and Duplicated Code~\cite{lacerda2020code}.

To confirm that CodeScene's Code Health adds predictive value beyond LoC, we analyzed its correlation with issue resolution times for our data set. Analogous to LoC counterpart, we calculated the Pearson correlation between the raw Code Health scores (10.0 down to 1.0) and the average issue resolution times in corresponding files. We found a higher degree of correlation (r=-0.58) and conclude that Code Health is a valid metric for this study.

\subsection{Measuring Time-in-Development} \label{sec:devtime}
In this study, we will measure the time related to resolving Jira issues and the subset of Jira issues referred to as defects (often called bugs). Note that issues encompass tasks beyond defects, e.g., feature requests and other types of non-corrective software maintenance. Jira distinguishes between two types of times during issue resolution. First, \textit{cycle time} is the time from the beginning to the end of a certain action, i.e., the time during which the issue has the status ``In progress''. Second, \textit{lead time} covers the entire time from receiving a request for an action to the moment this action is completed, i.e., including the time in the queue. 

\begin{figure*}
    \centering
    \includegraphics[width=0.9\textwidth]{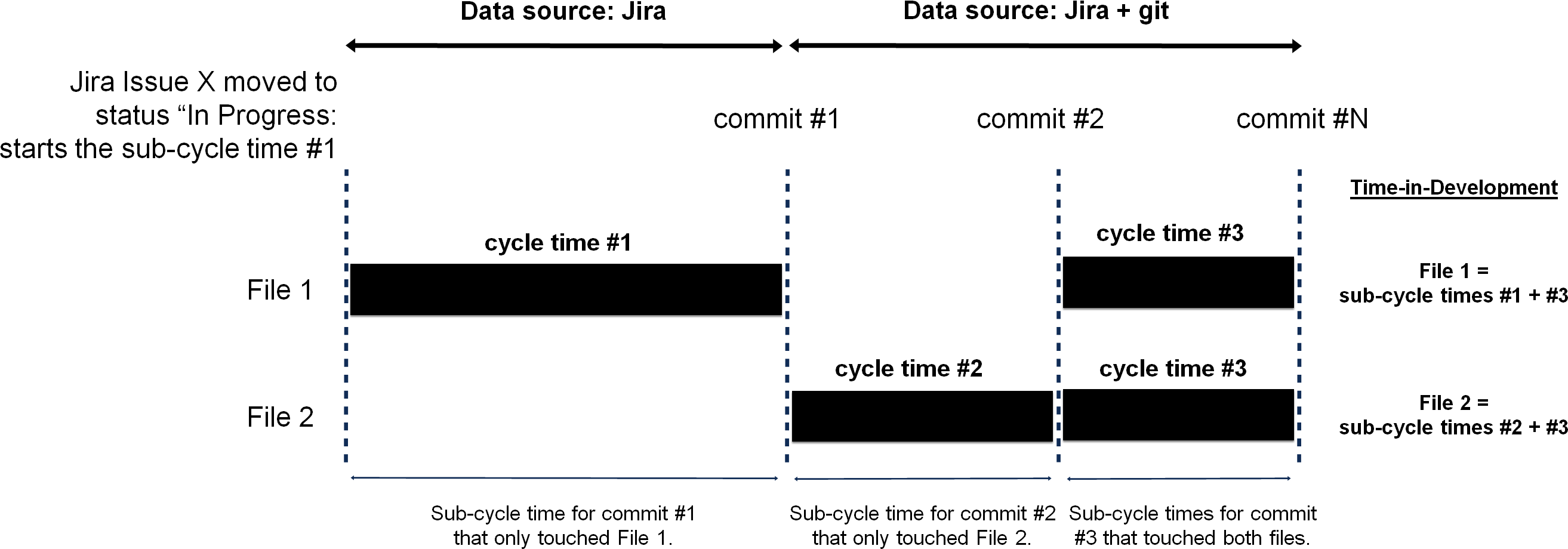}
    \caption{Overview of the Time-in-Development algorithm.}
    \label{fig:timeindevelopmentalgo}
\end{figure*}

We measure Time-in-Development as the cycle time spent on development per file so that we can correlate it with the Code Health on the file-level. While existing plugins for tools like Jira can measure the overall cycle time of an issue\footnote{\url{https://community.atlassian.com/t5/Marketplace-Apps-Integrations/3-tools-to-analyze-Cycle-and-Lead-Time-in-the-Jira/ba-p/1403775}}, the information on how that overall time is broken down to files of various quality is -- to the best of our knowledge -- not available in any existing Jira plugins. Hence, we decide to measure Time-in-Development using an algorithm that combines Jira status transition times with commit metadata from version control systems. As illustrated in Figure~\ref{fig:timeindevelopmentalgo}, we calculate the cycle time on a file level using the following algorithm: 

\begin{enumerate}
\item A developer starts work on a new feature, Jira Issue~X. Jira Issue~X requires changes to File~1. The cycle time for File~1 starts when the developer moves Jira Issue~X to the ``In~Progress'' status.
\item The cycle time for File~1 ends with the last commit referencing Jira Issue~X. If the developer considers the whole Jira issue implemented by now, i.e., moves it to the ``Done'' or ``Completed'' status, the Time-in-Development for Jira Issue~X would be this initial cycle time.
\item However, if Jira Issue~X corresponds to multiple commits, as indicated in Figure~\ref{fig:timeindevelopmentalgo}, then the time of the previous commit referencing Jira Issue~X is used to determine a sub-cycle time. This sub-cycle time is the time between commit~\#1 and commit~\#2.
\item The sub-cycle times for Jira Issue~X are summed up per file. This provides a Time-in-Development per file, independent of how many commits that were used to implement the Jira issue.
\item The preceding steps are performed for each completed Jira Issue, and the total Time-in-Development is accumulated per file.
\end{enumerate}

Our primary research scope is the actual time spent working on the code. Hence, the output of the preceding algorithm is the accumulated Time-in-Development per file for closed Jira issues. In contrast to many previous studies, we focus on aggregating cycle times rather than considering issue lead times. Focusing on the development stage eliminates other variables like company-specific post-development steps for deployment, releases, and testing practices.

\section{Related Work} \label{sec:rw}
\begin{figure*}
    \centering
    \includegraphics[width=0.8\textwidth]{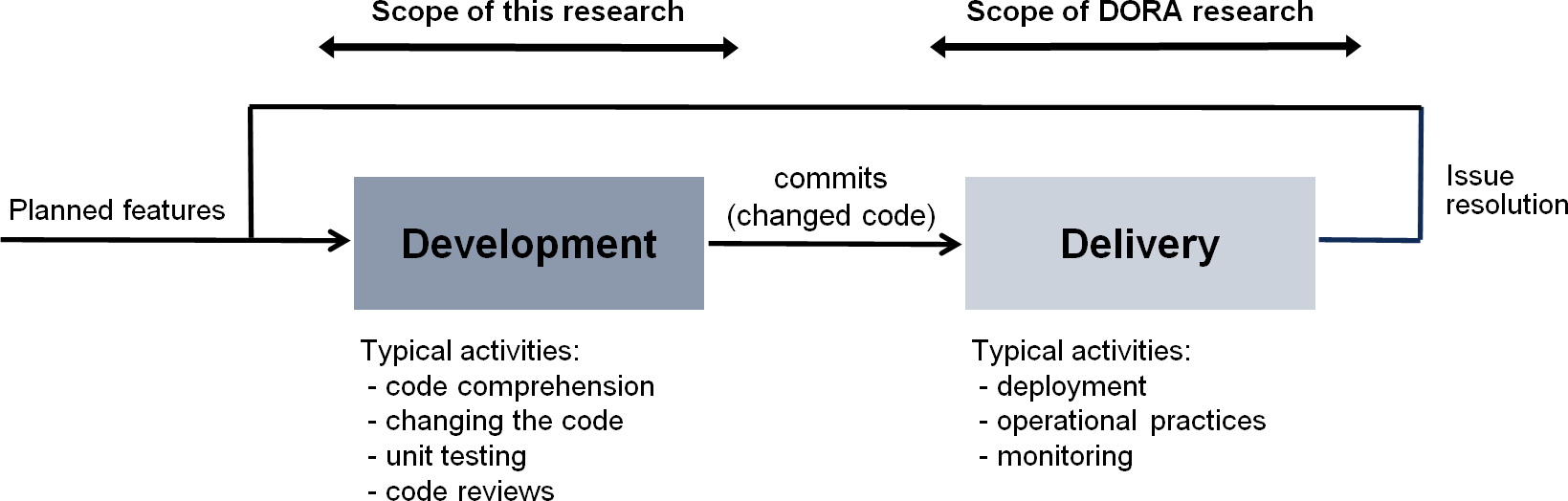}
    \caption{The scope of our research and how it complements DORA~\cite{humble2018accelerate}.}
    \label{fig:contextscopeofresearch}
\end{figure*}

There is no consensus of how to measure productivity in software engineering~\cite{oliveira2017have}. However, a promising recent addition is the related work by DevOps Research \& Assessment (DORA)\footnote{https://www.devops-research.com/research.html} that establishes productivity measures via its Four Key Metrics (FKM)~\cite{humble2018accelerate}. The FKM are 1) lead time for change, 2) deployment frequency, 3) time to restore service, and 4) change failure rate. These metrics are now becoming popular in the industry and receive much attention~\cite{sallin2021measuring}. Nonetheless, it is worth pointing out that the FKMs focus on the delivery side of software development, not the actual source code development. That is why our study aims to complement DORA's work by focusing on the earlier steps in the software development cycle: the waste that is introduced when the code is written as shown in Figure~\ref{fig:contextscopeofresearch}.

The research from the TD community indicates that this waste can be significant. Besker \textit{et al.} explored the consequences of TD~\cite{besker2018technical}. Their study was a longitudinal examination of software developers reporting their wasted time, and found that developers waste, on average, 23\% of their time due to TD. A white paper by Stripe reports that the number might be as high as 42\%~\cite{stripe}. Further, recent research on program comprehension points out that developers ``spend a long time on program comprehension just because the code quality is low''~\cite{xia2017measuring}. As such, any improvement in TD reduction or ease of understanding existing code would be an important advance for the industry.

There are existing methods for TD identification like the Maintainability Index, SIG TD models, and SQALE. These methods are based on static analysis techniques and code coverage measures~\cite{strevcansky2020comparing}. While these methods deliver quantitative metrics on TD, they lack a measure of the actual business impact in terms of the additional waste due to the identified TD. The cost of TD is not the time it would take to fix the code -- the remediation work -- but rather the additional development work due to technical quality issues~\cite{tornhill2018software}. That is, the interest an organization has to pay on its TD~\cite{nugroho2011empirical}.

Part of the TD interest might manifest itself as additional defects due to the lower quality of the code~\cite{snipes2012defining}. However, research on self-admitted TD has failed to find a clear trend when it comes to defects~\cite{wehaibi2016examining}. By applying the Code Health metric, this study aims to establish a more clear link between TD and defects.

Prior research by Lenarduzzi \textit{et al.} have calculated the variation in lead time for resolving Jira issues as an initial data-driven approach for TD estimation and to improve the prioritization when deciding to remove TD~\cite{lenarduzzi2021technical}. In their study, Lenarduzzi \textit{et al.} measured code TD using SonarQuebe's coding rules\footnote{\url{https://docs.sonarqube.org/latest/user-guide/rules/}}. They conclude that ``we have not found evidence that either the presence of a single violation or a number of violations affects the lead time for resolving
issues.'' In our study, we expect to identify that supporting evidence for finally establishing a link between TD and cycle times (i.e., Time-in-Development rather than lead times); since a prior student project reports that CodeScene's metric gives more significant results than the lower level issues reported by SonarQube~\cite{victoria}, our hypothesis is that our choice of code quality measure will have a significant effect on the lead times. The Code Health metric itself has been found to predict security vulnerabilities~\cite{al2021presence}, but, to the best of our knowledge, no prior research has evaluated the relationship between code quality on a file level and the cycle Time-in-Development. With our study, we aim to provide numbers that present the wasted time in general terms, allowing development organizations to put a value on TD reducing activities and code quality in general.

Several previous software engineering studies have investigated the business side of technical debt. Ampatzoglou et al. conducted a systematic literature review on TD with a focus on financial aspects~\cite{ampatzoglou2015financial}. Based on a selection of 69 primary studies, they identified the use of several financial approaches to TD management, e.g., portfolio management, cost/benefit analysis, and value-based analysis. However, the authors found that these approaches have been inconsistently applied and called for more research on closing the gap between business and engineering departments. To help align the two perspectives, the authors developed a glossary and a classification scheme. Our work also aims at bridging business and engineering departments, but this study focuses on providing empirical evidence on the costs of TD, allowing development organizations to put a value on TD reducing activities and code quality in general.

More recent work, i.e., published after Ampatzoglou \textit{et al.}’s review~\cite{ampatzoglou2015financial}, includes work by de~Almeida and colleagues and Besker \textit{et al.} Almeida et al. developed Tracy, a ``business-oriented decision-making framework to promote the alignment between technical decisions and business expectations’’~\cite{de2019tracy}. The development followed the design science research paradigm and involved 49 professionals from three companies. TD management is at the heart of Tracy, and lessons learned from an industrial case study indicate that the framework can help developers focus on the TD that matters the most~\cite{de2021business}. We believe that the Code Health measurements in this paper could enter the high-level Tracy framework and make its resulting TD prioritization more accurate. Finally, Besker et al. estimated wasted time caused by the TD interest during the software life-cycle~\cite{besker2017pricey}. Based on survey research and 32 in-depth interviews, the authors found that, on average, 36\% of all development time is wasted due to TD. Moreover, Besker et al. concluded that most of the time is wasted on understanding and measuring TD. We believe that the automated Code Health measurements studied in this paper could remedy parts of this waste.

%With our study, we aim to provide numbers that present the wasted time in general terms, allowing development organizations to put a value on TD reducing activities and code quality in general.

\section{Method} \label{sec:method}
This study follows the draft version of the empirical standard for repository mining\footnote{\url{https://github.com/acmsigsoft/EmpiricalStandards/blob/master/docs/RepositoryMining.md} (2021-12-10, Latest commit 8c65057)}. Repository mining is the appropriate standard as we use automated techniques to extract data from source code repositories and issue management systems followed by quantitative analysis. The selected repositories represent 39 proprietary software development projects that use CodeScene for source code analysis. For confidentiality reasons, we cannot disclose neither the selection criteria nor the procedure for acquisition -- but we stress that all repository owners have provided their consent to let us analyze their data for this study. Regarding metrics, our analysis relies on established software engineering measures complemented by two additional metrics: Code Health and Time-in-Development described in Sections~\ref{sec:codehealth} and~\ref{sec:devtime}, respectively.

\subsection{Research Questions}
The research goal of this study is to evaluate the impact of code quality on defect density and the time spent in development, i.e., cycle times. Our overall purpose is to raise awareness of code quality in the industry by providing evidence that code quality needs to be a business concern, reported and attended to with the same rigor as financial KPIs. Based on this goal and purpose, this study examines three research questions:

\begin{itemize}
    \item[RQ1] How does the number of reported defects in source code files correlate to source code quality?
    \item[RQ2] How much longer development time is needed to resolve an issue in files with low quality source code?
    \item[RQ3] To what extent is the code quality of a file related to the predictability of resolving issues on time?
\end{itemize}

The rationales behind our RQs are as follows. RQ1 investigates the number of reported Jira defects to explore whether low quality code is more bug prone. More defects would, in TD terms, translate into a higher interest due to the increased maintenance work. RQ2 gets to the core of our research aims and investigates if resolving Jira issues in low quality code takes longer than comparable changes to high quality code. Finally, the objective of RQ3 is to examine if resolving issues in low quality code involves a higher risk -- in terms of increased maximum issue resolution times in such files compared to high quality code. 

\subsection{Data Collection} \label{sec:data}
The data for the study was mined via the CodeScene tool, which is available for free to the research community. A selection of CodeScene's customers were asked to participate in this study, allowing us to tap into CodeScene's data lake consisting of code metrics, trends, and process-related data. All data owners have given their consent for us to mine their repositories for the purpose of this study, given that the companies remain anonymous and that the data cannot be tracked back to their specific codebase, product, or business entity.

\begin{figure}
    \centering
    \includegraphics[width=0.45\textwidth]{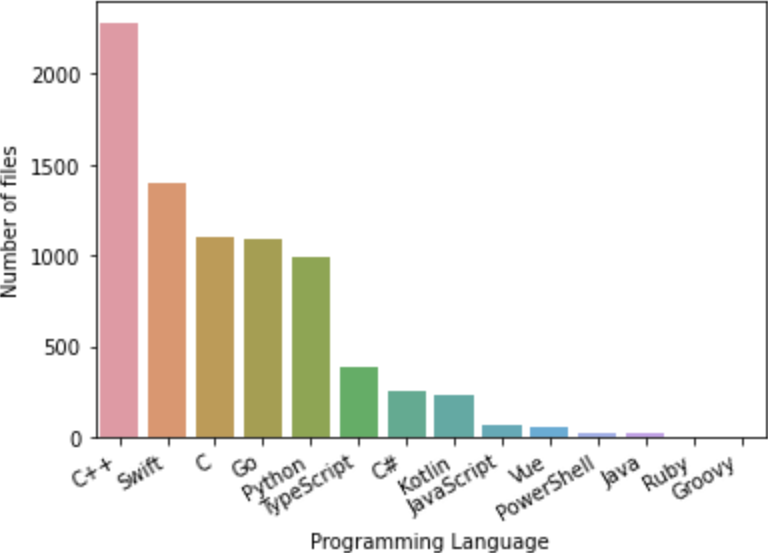}
    \caption{Overview of the studied programming languages.}
    \label{fig:programminglanguages}
\end{figure}

In total, we collected data from 39 proprietary codebases under active development. Table~\ref{tab:desc_stats} shows non-sensitive summary statistics for the 39 projects. The collected data represents development tasks completed over 6-12 months, depending on project. The codebases are from several different industry segments, including retailing, construction, infrastructure, brokers, data analysis, and more. The codebases are implemented in various programming languages as illustrated in Figure~\ref{fig:programminglanguages}. The resulting dataset contains 30,737 source code files for which we extracted Code Health scores and values for Time-in-Development. A replication package, containing anonymized data and a computational notebook, is available on GitHub\footnote{\url{https://github.com/empear-analytics/code-health-study-tech-debt-2022}}.

\begin{table}[]
\caption{Overview of the studied projects.}
\label{tab:desc_stats}
\begin{tabular}{lccc|}
\cline{2-4}
\multicolumn{1}{l|}{}                   & Average      & 75\%      & Std        \\ \hline
\multicolumn{1}{|l|}{kLoc}              & 171.42       & 155.0     & 524.60     \\
\multicolumn{1}{|l|}{Avg. \#issue/file} & 4.17         & 4.0       & 7.13       \\ \hline
\multicolumn{4}{|c|}{Distribution of Code Health}                               \\ \hline
\multicolumn{1}{|l|}{\#Healthy files}   & \multicolumn{3}{c|}{27,352 (88.99\%)} \\
\multicolumn{1}{|l|}{\#Warning files}   & \multicolumn{3}{c|}{3,035 (9.87\%)}   \\
\multicolumn{1}{|l|}{\#Alert files}     & \multicolumn{3}{c|}{350 (1.14\%)}     \\ \hline
\end{tabular}

\end{table}

Since Jira and version-control data can be noisy~\cite{aranda2009secret}, we filtered the data with respect to outliers for Time-in-Development. Outliers were defined as data points that are more than 1.5 * IQR (interquartile range) above the third quartile or below the first quartile.

\subsection{Analysis and Synthesis}
\begin{figure}
    \centering
    \includegraphics[width=0.5\textwidth]{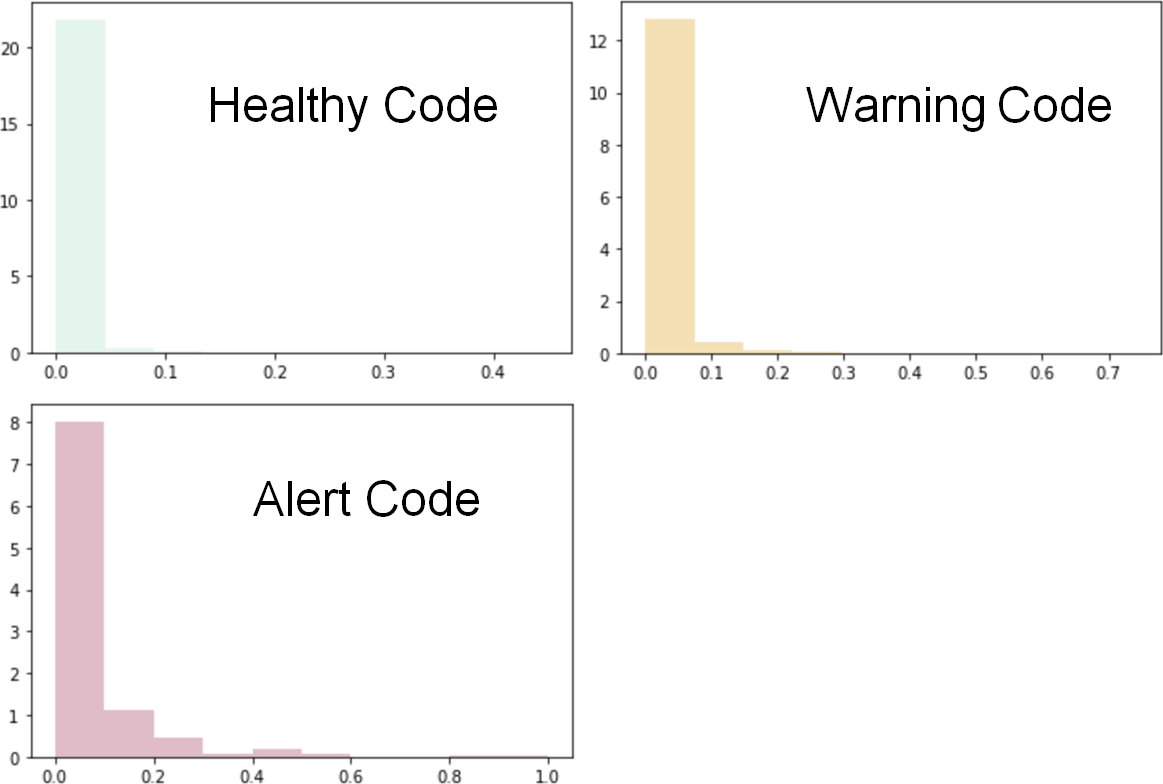}
    \caption{Jira defects for DV-RQ1 by Code Health category. Y-axes show probability densities and X-axes represent the scaled number of defects per file.}
    \label{fig:jiradefectsdistribution}
\end{figure}

The distribution of Jira defects per file studied in RQ1 does not meet the requirements for normality of the Dependent Variables (DV). Looking at the distributions shown in Figure~\ref{fig:jiradefectsdistribution}, i.e., DV-RQ1, reveals that the data is positively skewed. This is because most files didn't have any fixed defects during the studied time period. Removing those lower outliers would bias the results as such a step would remove an important part of what is being studied: defect free code is a strong signal.

To account for the non-normality of the data distribution together with our large sample size, we apply the Kruskal-Wallis nonparametric statistical test to assess the differences between the Code Health categories~\cite{ostertagova2014methodology}. We perform post-hoc comparisons between the different Code Health categories using the Dwass, Steel, Critchlow and Fligner (DSCF) all-pairs comparison test~\cite{neuhauser2001nonparametric}. 

For Time-in-Development, we conducted two one-way analyses of variance (ANOVA) where the independent variable is the Code Health category (Healthy, Warning, or Alert), and the dependent variable varies depending on the RQ:
\begin{itemize}
    \item[DV-RQ2] the average Time-in-Development to resolve a Jira issue per file
    \item[DV-RQ3] the average maximum Time-in-Development for resolving a Jira issue per file.
\end{itemize}

\begin{figure}
    \centering
    \includegraphics[width=0.49\textwidth]{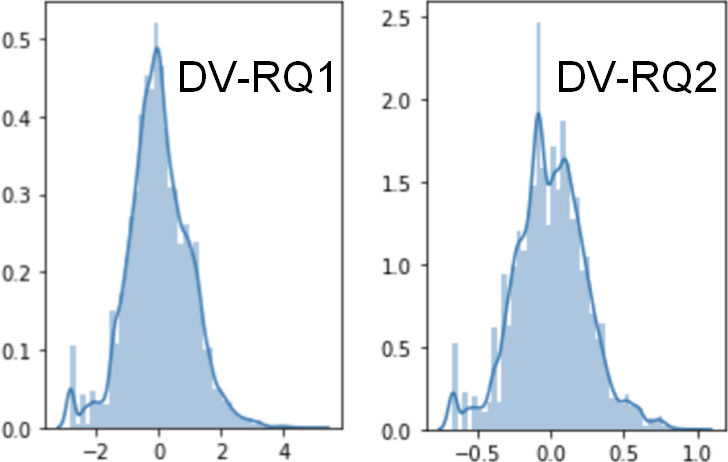}
    \caption{Normalized DV-RQ2 and DV-RQ3 for ANOVA.}
    \label{fig:normalizedanovadata}
\end{figure}

To meet the pre-conditions for an ANOVA~\cite{cardinal2013anova}, we must ensure normality of the dependent variables. We normalize DV-RQ2 and DV-RQ3 using the Yeo-Johnson algorithm~\cite{yeo2000new}. Figure~\ref{fig:normalizedanovadata} shows the result.

Finally, we performed post-hoc comparisons between the different Code Health categories using Tukey's HSD test~\cite{tukey1957comparative}. The correlation of the Code Health categories with the dependent variables were also analyzed visually based on bar charts with plotted error bars.

\section{Results} \label{sec:results}
The following subsections present the results for the RQs.

\subsection{Code Quality and Number of Issues (RQ1)}
\begin{figure}
    \centering
    \includegraphics[width=0.5\textwidth]{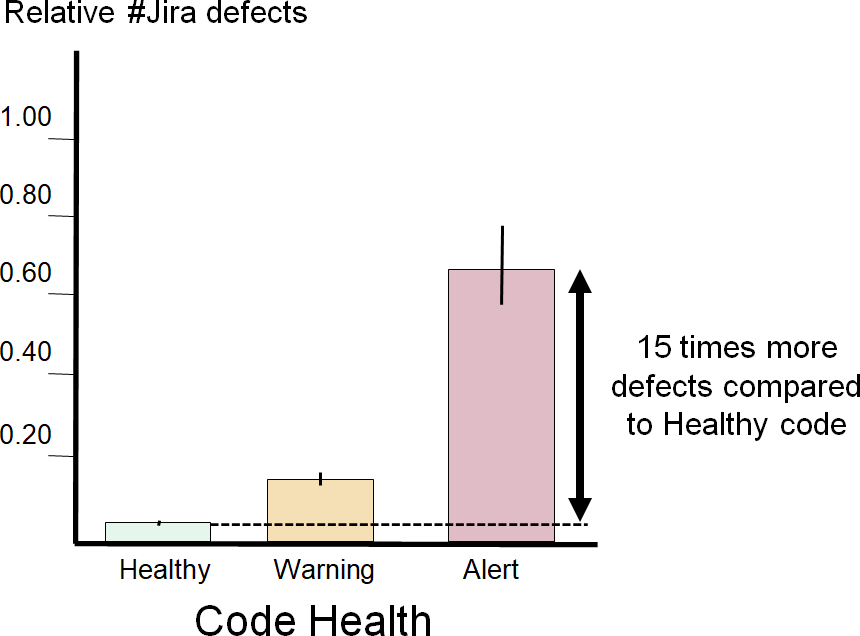}
    \caption{Average number of Jira defects per file for each category (scaled). The standard errors are presented as vertical lines.}
    \label{fig:numberofissues}
\end{figure}

Table~\ref{tab:resultsissues} presents the average number of Jira defects for the three Code Health categories. Figure~\ref{fig:numberofissues} gives a graphical rendition of the scaled (0.0 to 1.0) average number of Jira defects for the three Code Health categories. The scaled averages represent for Healthy, Warning, and Alert are 0.004, 0.016, and 0.06, respectively. A Kruskal-Wallis test shows that there is a statistically significant difference in the number of Jira defects between the different Code Health categories, $x2(2) = 1645.05, p = 0.0$.

\begin{table}[]
\caption{Average number of Jira defects per file for each Code Health category.}
\label{tab:resultsissues}
\begin{tabular}{lc|ccc|c}
                                 &         & Healthy  & Warning  & Alert  & All      \\ \hline
\multirow{3}{*}{Jira defects}   & Avg & 0.25 & 0.94 & 3.70 & 0.35 \\
                                 & 75\%    & 0.00   & 1.00   & 4.00   & 0.0   \\
                                 & Std     & 0.90   & 2.58   & 6.61   & 1.43   \\ \hline
\end{tabular}
\end{table}

The DSCF test for multiple comparisons found that the average number of Jira defects was significantly different between the Healthy and Warning categories ($p = 0.001$), between the Warning and Alert categories ($p = 0.001$), as well as between the Healthy and Alert categories ($p = 0.001$). There is a medium to large effect size in the differences between Healthy and Alert with respect to the number of Jira defects (Cohen's $d=0.73$).

Based on these results, we find that code quality has a significant impact on the number of defects in a file (RQ1). Comparing the averages suggests that Alert code results in 15 times more reported Jira defects than Healthy code.

\subsection{Code Quality and Time-in-Development (RQ2)}
\begin{figure}
    \centering
    \includegraphics[width=0.5\textwidth]{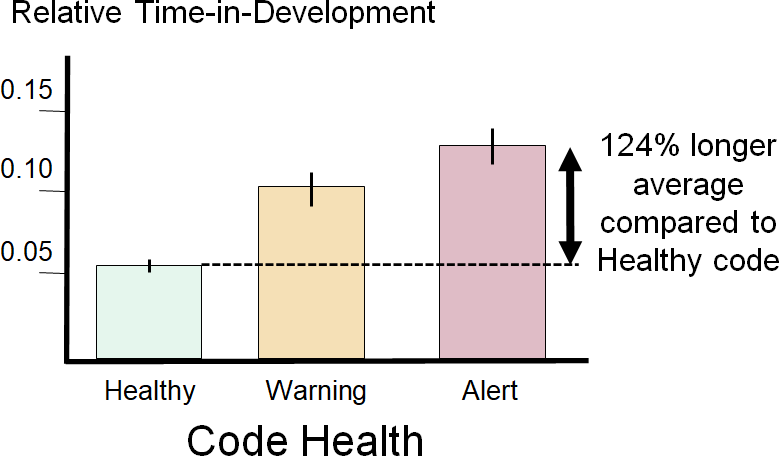}
    \caption{Average Time-in-Development (scaled) for resolving a Jira issue per file. The standard errors are depicted as vertical lines.}
    \label{fig:meantimeforimplementation}
\end{figure}

The upper part of Table~\ref{tab:results} presents the average Time-in-Development, in minutes, corresponding to the three Code Health categories. A graphical rendition of the scaled (0.0 to 1.0) average values is given in Figure~\ref{fig:meantimeforimplementation}. The normalized averages for Healthy, Warning, and Alert are -0.05, 0.33, and 0.85, respectively. These are significantly different, $F(2, 30734) = 742.56, p = 0.00$.

\begin{table}[]
\caption{Time-in-Development per file for each Code Heath category (minutes).}
\label{tab:results}
\begin{tabular}{cl|ccc|c}
                                                                                          &      & Healthy & Warning & Alert    & All     \\ \hline
\multirow{3}{*}{\begin{tabular}[c]{@{}c@{}}Time-in-\\ Development\end{tabular}}           & Avg  & 7815.6  & 13934.6 & 17544.5  & 8573.1  \\
                                                                                          & 75\% & 7320.0  & 12165.0 & 21661.5  & 8014.5  \\
                                                                                          & Std  & 22405.8 & 43162.9 & 20630.1  & 25392.5 \\ \hline
\multirow{3}{*}{\begin{tabular}[c]{@{}c@{}}Maximum\\ Time-in-\\ Development\end{tabular}} & Avg  & 15111.9 & 34024.5 & 129940.3 & 18286.9 \\
                                                                                          & 75\% & 14040.0 & 30900.0 & 184320.0 & 16260.0 \\
                                                                                          & Std  & 37719.1 & 78253.8 & 164057.2 & 48492.4 \\ \hline
\end{tabular}
%\begin{tabular}{lc|ccc|c}
%                                 &         & Healthy  & Warning  & Alert  & All      %\\ \hline
%\multirow{3}{*}{Time}   & Avg & 7815.6 & 13934.6 & 17544.5 & 8573.1 \\
%                                 & 75\%    & 7320.0   & 12165.0   & 21661.5   & 8014.5   \\
%                                 & Std     & 22405.8   & 43162.9   & 20630.1   & 25392.5   \\ \hline
%\multirow{3}{*}{Maximum time}   & Avg & 15111.9 & 34024.5 & 129940.3 & 18286.9 \\
%                                 & 75\%    & 14040.0   & 30900.0   & 184320.0   & 16260.0   \\
%                                 & Std     &  37719.1  & 78253.8   & 164057.2   & 48492.4   \\ \hline
%\end{tabular}
\end{table}

Tukey’s HSD Test for multiple comparisons found that the average Time-in-Development was significantly different between Healthy and Warning categories ($p = 0.001$), between the Warning and Alert categories ($p = 0.001$), as well as between the Healthy and Alert categories ($p = 0.001$). There is a small to medium effect size in the differences between Healthy and Alert for the average Time-in-Development (d=0.45).

Based on these results, we find that code quality has a significant impact on Time-in-Development (RQ2). Comparing the scaled average Time-in-Development gives that, on average, development needs 124\% longer in the Alert category compared to development time in Healthy code. Further, we note that already in the Warning category, the average development time for a Jira issue is 78\% longer than in Healthy code.

\subsection{Code Quality and predictable task completion (RQ3)}
\begin{figure}
    \centering
    \includegraphics[width=0.5\textwidth]{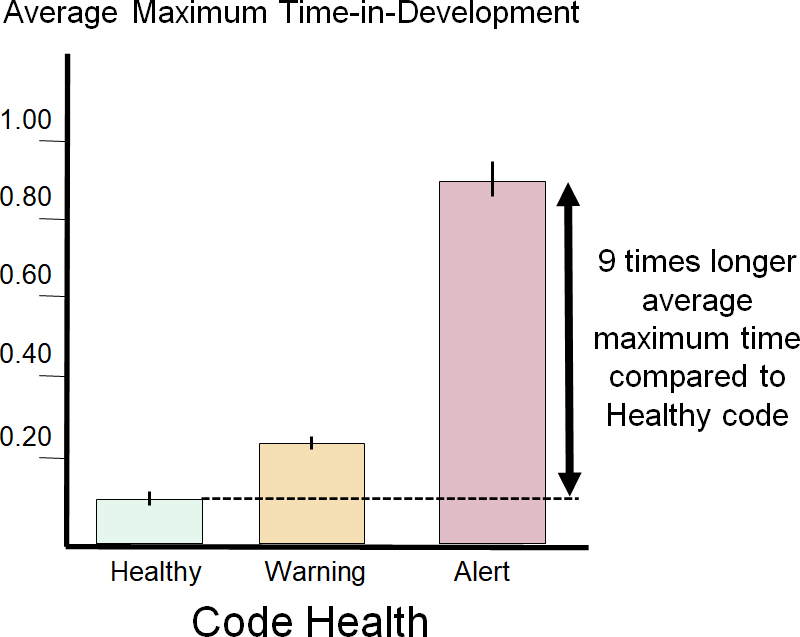}
    \caption{Average maximum Time-in-Development (scaled) for resolving a Jira issue per file. The standard errors are shown as vertical lines.}
    \label{fig:maxtimeforimplementation}
\end{figure}

The lower part of Table~\ref{tab:results} presents the average maximum Time-in-Development for files corresponding to the three Code Health categories. Figure~\ref{fig:maxtimeforimplementation} depicts a graphical rendition of the scaled (0.0 to 1.0) average values. The normalized averages for the maximum Time-in-Development per file for Healthy, Warning, and Alert are -0.02, 0.10, and 0.36, respectively. These are significantly different, $F(2, 30734) = 1399.55, p = 0.00$.

Tukey’s HSD test for multiple comparisons found that the average value of the maximum Time-in-Development for a Jira issue was significantly different between Healthy and Warning categories ($p = 0.001$), between Warning and Alert categories ($p = 0.001$), as well as between Healthy and Alert categories ($p = 0.001$). There is a large effect size in the differences between Healthy and Alert for the maximum Time-in-Development (Cohen's $d=0.96$).

Based on these results, we find that the Time-in-Development in low quality code is significantly less predictable in the sense that the maximum development time for a Jira issue increases with lower code quality. Comparing the scaled averages shows that the maximum Time-in-Development for the Alert category is 9 times longer than the maximum Time-in-Development for Healthy code. Further, we note that already in the Warning category, the average maximum Time-in-Development is more than twice as long compared to Healthy code.

\subsection{Threats to Validity}
As any empirical study, our work is subject to threats to validity. We discuss these threats organized into internal validity, external validity, construct validity, and reliability~\cite{runeson2012case}.

The major threat to the internal validity of this research is the relationship between the dependent variables number of defect (DV-RQ1) and Time-in-Development (DV-RQ2 and DV-RQ3) on one hand, and the Code Health categories on the other. Software development is a complex endeavor, meaning there can be confounding variables such as differences in process~\cite{deissenboeck2007economic} and organizational factors~\cite{nagappan2008influence}. To mitigate this threat, we included a wide variety of codebases from a diverse set of organizations. This study is also restricted to the development phase of the software, meaning that differences in post-development activities such as deployment are not included in the Time-to-Development measurements, i.e., we focus on Jira cycle time. We also acknowledge the need for further studies to gain additional insights into the potentially confounding variables in Section~\ref{sec:fw}. 

Another internal validity threat relates to our causality claims. Does low quality code lead to additional defects or is the code quality low because of the many defects? In the same vein, although less probable, code quality could be lower because developers resolve issues in corresponding files too slowly -- they might be less attentive when working under stronger perceived time pressure and thus produce lower quality code. We accept these threats and highlight that a future investigation of causality could rely on randomized controlled trials or Bayesian causal analysis~\cite{hernan2020causal}. Finally, the differences in sample size of the different Code Health groups might be a threat to the internal validity: there is simply more high quality than low quality code in our data set (cf. Table~\ref{tab:desc_stats}).

The external validity refers to the extent to which the findings are possible to generalize to software development in general. All included codebases come from CodeScene users, which might be a sample of companies that are not representative of the entire software development landscape. To mitigate this risk, our study relies on a large number of participating organizations from different domains and business as specified in Section~\ref{sec:data}. A further validity threat might be differences in the number of defects due to the programming language of choice. Even though our data set includes samples from 14 different programming languages, the main language is C++. From a validity perspective, we note that prior research on the impact of programming languages on code quality has failed to find a causation~\cite{berger2019impact}.

Construct validity refers to the extent to which the studied measures represent what the researchers are considering. To mitigate the threat to the construct validity of Code Health, we performed a Pearson correlation study to ensure that Code Health adds predictive value beyond LoC (see Section~\ref{sec:codehealth}). As mentioned in Section~\ref{sec:rw}, the Code Health metric has also been shown to find more significant issues than SonarQube -- a widely adopted tool used by 200K development teams\footnote{https://www.sonarqube.org/} -- as well as predicting security vulnerabilities~\cite{al2021presence}.

The Time-in-Development construct is based on a combination of Jira and version-control data. As such, a threat to the construct validity is that Time-in-Development is constrained by what those two data sources capture. One such validity threat is that a git commit can -- and often do -- contain multiple modified files. Using the algorithm described in Section~\ref{sec:devtime}, all of these files will get the same Time-in-Development even though the actual development time might be shorter for any of the files in the commit. We aimed to reduce this threat by collecting a large set of files (40.000).

Reliability refers to the goal of minimizing the errors and biases in a study. Reliability means that the study is replicable by other researchers that follow the same procedures as the original study. To mitigate reliability threats, our study follows the evolving community standard for software engineering studies that use automated techniques to extract data from large-scale data repositories~\cite{ralph2020empirical}.

\section{Discussion} \label{sec:disc}
There are multiple reasons why development organizations take on TD in the shape of low quality code\footnote{https://martinfowler.com/bliki/TechnicalDebtQuadrant.html}. As pointed out in the introduction, this type of waste accounts for up to 42\% of developers' time. Given those numbers, it is surprising that as few as 7.2\% of organizations methodically track TD~\cite{martini2018technical}. Further, a recent study from the video game industry shows that while there is high awareness of TD, paying off the accumulated debt is a rare practice~\cite{borowa2021living}.
One explanation for this disconnect between awareness and actions is that little is known about the effects of TD and TD payment practices~\cite{freire2021technical}. In our research, we provide hard numbers on the value of code quality in order to raise the awareness of the costs of TD as well as providing an expected return on investment if TD is paid off.

The results from RQ1 shows that low quality code in the Alert category has up to 15 times more reported Jira defects than Healthy code. This is important from a business perspective as the cost of poor quality software in the US alone costs approximately \$2.84 trillion per year~\cite{krasner2018cost}. Software defects also translate into unplanned work, which leads to cost overruns~\cite{nonaka2007project}. Finally, there is an opportunity cost that comes with software defects: when a significant amount of development time is spent fixing defects, less time remains for implementing new features that keep the product competitive~\cite{snipes2012defining}.

In RQ2 we investigated the additional Time-in-Development for low quality code. Our results show that resolving a Jira issue in code in the Alert category needs, on average, 124\% longer than in Healthy code. These results should have a dramatic impact on how organizations value code quality as it translates into a time to market that is more than twice as long as it needs to be. Given that companies must hire high-skilled developers from an increasingly shrinking labor supply, wasting precious time on low quality code puts the business at a disadvantage~\cite{breaux20212021}.

Finally, RQ3 investigated the predictability for resolving Jira issues. The results show that implementing a Jira issue in low quality code in the Alert category might require up to 9 times longer Time-in-Development compared to corresponding changes to Healthy code. In an industry setting, this increase in variance is likely to lead to lower predictability for task completion -- accurate effort estimation is indeed an acknowledged challenge that has been researched for decades~\cite{kocaguneli2011value}. Moreover, low predictability would translate into increased irritation and stress levels of managers~\cite{mohr2010stress}. Hence, the effects of low code quality go beyond a purely financial impact as it can impact the whole team since one post stimulation effect of stress is an insensitivity toward others~\cite{cohen1980aftereffects}. This human link has implications for the future of our industry. As pointed out in Beyond Legacy Code, ``if we want to have a true industry around developing software then it has to be sustainable''~\cite{bernstein2015beyond}. The results of our study shows that low quality code is at odds with that goal; working on a codebase where tasks take an order of magnitude longer to resolve is anything but sustainable. Code quality matters if we want to build a sustainable software industry.

\section{Conclusions and future work} \label{sec:conc}
This research was initiated to make code quality a business concern by putting numbers on the impact of low quality code. We chose this line of research since code quality has been an abstract concept that rarely get adequate attention at the management level~\cite{martini2018technical}.

Our results show that there is a significant impact of code quality on both time to market as well as the external quality of the product in the shape of software defects. While the impact was strongest in the ``Red'' Code Health category, i.e., Alert code as provided by CodeScene, with 124\% longer development time, already Warning Code Health needed 78\% longer to implement a Jira task than in Healthy code. This means that even codebases with a moderate degree of code quality issues pay a significant price.

We consider these results of topmost importance to the software industry: due to the predicted global shortage of software developers~\cite{breaux20212021}, organizations cannot hire as many developers as the perceived need. Our results indicate that improving code quality could free existing capacity; with 15 times fewer bugs, twice the development speed, and substantially more predictable issue resolution times, the business advantage of code quality should be unmistakably clear.

\subsection{Future Work} \label{sec:fw}
During this research we identified a number of directions for future studies that would help in investigating potentially confounding variables as well as expand the knowledge base with respect to how code quality impacts software development.

The results for RQ3 were dramatic, and we plan further studies to deepen the knowledge on why changes to low quality code have such a high variance and uncertainty. One interesting research path is the knowledge dimension. A file with production code typically has a main author and possible minor contributors that have made smaller contribution to the overall file. Could it be that changes done by the main author are faster and more predictable than corresponding changes made by a minor contributor? We suspect that author experience could be a factor that impacts the predictability of changes to low quality code. Some research seem to suggest a link: ``We also find evidence of the mediating effect of knowledge on the effects of time pressure''~\cite{kuutila2020time}. If such a relationship exists, it would help the industry by highlighting the organizational risks -- such as key personnel dependencies -- associated with low quality code.

Our current study includes codebases written in many (14) different programming languages. This data set would also allow us to study potential differences for Time-in-Development depending on the programming language of choice. Such a difference is claimed in ``Software Estimation: Demystifying the Black Art'' by McConnell~\cite{mcconnell2006software}, and a future study would present up to date data if such a relationship exists.

Finally, our current data set for Time-in-Development contains both Jira defects and other types of issues, e.g., feature requests and improvements. We plan to conduct a follow-up study that separates Time-in-Development for planned work (features, improvements) vs. unplanned work (defect resolution). Such a study would offer additional insights into why the unpredictability is so much higher in low quality code. A working hypothesis is that debugging is much more time-consuming source good of low quality.

\section*{Acknowledgements}
Our thanks go to the CodeScene development team who supported our work and provided details on the Code Health metric. Moreover, we extend our deepest appreciation to the repository owners who let us analyze their data as part of this study.